\documentclass[letter]{aa}
\usepackage{natbib,twoopt}
\usepackage{xcolor}
\usepackage[hyphenbreaks]{breakurl}
\usepackage[breaklinks]{hyperref}      %% to avoid \citeads line fills, add "draft" 
\bibpunct{(}{)}{;}{a}{}{,}             %% natbib format for A&A and ApJ
\definecolor{cobalt}{rgb}{0.06, 0.2, 0.65}
\hypersetup{
  colorlinks,
  citecolor=cobalt,
  linkcolor=[rgb]{0.8, 0.2, 1.0},
  urlcolor=cobalt,
}
\makeatletter
  \newcommandtwoopt{\citeads}[3][][]{\href{http://adsabs.harvard.edu/abs/#3}%
    {\def\hyper@linkstart##1##2{}%
     \let\hyper@linkend\@empty\citealp[#1][#2]{#3}}}
  \newcommandtwoopt{\citepads}[3][][]{\href{http://adsabs.harvard.edu/abs/#3}%
    {\def\hyper@linkstart##1##2{}%
     \let\hyper@linkend\@empty\citep[#1][#2]{#3}}}
  \newcommandtwoopt{\citetads}[3][][]{\href{http://adsabs.harvard.edu/abs/#3}%
    {\def\hyper@linkstart##1##2{}%
     \let\hyper@linkend\@empty\citet[#1][#2]{#3}}}
  \newcommandtwoopt{\citeyearads}[3][][]%
    {\href{http://adsabs.harvard.edu/abs/#3}
    {\def\hyper@linkstart##1##2{}%
     \let\hyper@linkend\@empty\citeyear[#1][#2]{#3}}}
\makeatother

\usepackage{txfonts}
\usepackage{hyperref}
\usepackage{upgreek}
\usepackage{newtxtext,newtxmath}

\usepackage[T1]{fontenc}

\usepackage{graphicx}	% Including figure files
\usepackage{amsmath}	% Advanced maths commands
\usepackage{amssymb}	% Extra maths symbols
\usepackage[normalem]{ulem}
\usepackage{caption}
\usepackage{subcaption}
\usepackage{multirow}
\usepackage{adjustbox} % Limit table
\usepackage{lscape}  %landscape formatting
\usepackage{array}
\usepackage{float}
\usepackage{orcidlink}

\newcommand{\msol}{\,M$_{\odot}$}
\newcommand{\sfr}{\,M$_{\odot}$\,yr$^{-1}$}
\newcommand{\sfrd}{\,M$_{\odot}$\,yr$^{-1}$\,kpc$^{-2}$}
\newcommand{\um}{\,$\upmu$m}

\newcolumntype{C}[1]{>{\centering\arraybackslash}p{#1}}

\begin{document}

\title{A nuclear spiral in a dusty star-forming galaxy at $z=2.78$}
\titlerunning{A nuclear spiral in a dusty star-forming galaxy at $z=2.8$}

\author{
\orcidlink{0000-0002-8999-9636}H.\,R.~Stacey\inst{1}\thanks{E-mail: hannah.stacey@eso.org},
\orcidlink{0000-0002-1173-2579}M.~Kaasinen\inst{1},
\orcidlink{0000-0003-2227-1998}C.\,M.~O'Riordan\inst{2},
\orcidlink{0000-0003-1787-9552}J.\,P.~McKean\inst{3,4,5},
\orcidlink{0000-0002-4912-9943}D.\,M.~Powell\inst{2}
\and
\orcidlink{0000-0001-9705-2461}F.~Rizzo\inst{3}
}
\authorrunning{H.\,R.~Stacey et al.}

\institute{
% List of institutions
European Southern Observatory, Karl-Schwarzschild Str. 2, D-85748 Garching bei M\"unchen, Germany 
\and
Max Planck Institute for Astrophysics, Karl-Schwarzschild Str. 1, D-85748 Garching bei M\"unchen, Germany
\and
Kapteyn Astronomical Institute, University of Groningen, P.O. Box 800, 9700 AV Groningen, The Netherlands
\and
South African Radio Astronomy Observatory (SARAO), P.O. Box 443, Krugersdorp 1740, South Africa
\and
Department of Physics, University of Pretoria, Lynnwood Road, Hatfield, Pretoria, 0083, South Africa
}

\date{Received 7 Oct 2024; accepted 4 Dec 2024}

\abstract{
The nuclear structure of dusty star-forming galaxies is largely unexplored but harbours critical information about their structural evolution. Here, we present long-baseline Atacama Large (sub-)Millimetre Array (ALMA) continuum observations of a gravitationally lensed dusty star-forming galaxy at $z=2.78$. We use a pixellated lens modelling analysis to reconstruct the rest-frame 230\um\ dust emission with a mean resolution of $\approx55$~pc and demonstrate that the inferred source properties are robust to changes in lens modelling methodology. The central 1 kpc is characterised by an exponential profile, a dual spiral arm morphology and an apparent super-Eddington compact central starburst. We find tentative evidence for a nuclear bar in the central 300~pc. These features may indicate that secular dynamical processes play a role in accumulating a high concentration of cold gas that fuels the rapid formation of a compact stellar spheroid and black hole accretion. We propose that the high spatial resolution provided by long-baseline ALMA observations and strong gravitational lensing will give key insights into the formation mechanisms of massive galaxies.
}


\maketitle

%%%%%%%%%%%%%%%%%%%%%%%%%%%%%%%%%%%%%%%%%%%%%%%%%%

%%%%%%%%%%%%%%%%% BODY OF PAPER %%%%%%%%%%%%%%%%%%

\section{Introduction}

At the peak of cosmic star formation (`cosmic noon'), dust-obscured star-forming galaxies (DSFGs) make up the high end of the stellar mass function ($M_\star \gtrsim 10^{10}$\msol) with star formation rates that can exceed $1000$\sfr\ \citep{Casey:2014,Hodge:2020}. Rapid gas accumulation and feedback from star formation and active galactic nuclei (AGN) are predicted to play important roles in the evolution of DSFGs into quiescent compact spheroids \citep{Hopkins:2009}. Yet many open questions remain about the structural evolution of such galaxies through cosmic time \citep{Conselice:2014}.

To form an in-situ stellar spheroid, gas must lose angular momentum such that it flows into the central kpc faster than it can turn into stars \citep{Dekel:2014}. It has long been predicted that gas-rich mergers play a critical role in both gas delivery and loss of angular momentum by inducing turbulence, either via major mergers \citep{Sanders:1996,Mihos:1996,Hopkins:2009} or fragmentation of gas discs due to gravitational instabilities induced by minor mergers and accretion \citep{Krumholz:2018,Kretschmer:2022}. However, recent work has revealed an abundance of rotationally supported (dynamically cold) discs in DSFGs at $z=2\to5$ with rotation-to-random motions comparable to local discs (e.g. \citealt{Lelli:2021,Lelli:2023,Rizzo:2021,Rizzo:2023,Fraternali:2021,Tsukui:2021,Liu:2024}). These galaxies have levels of turbulence that can be fully explained by stellar feedback alone, contrary to canonical model predictions \citep{Rizzo:2024}. Additionally, suggestions of spiral arms and bars in other $z=2\to4$ DSFGs on kpc-scales \citep{Hodge:2019,Costantin:2023,Amvrosiadis:2024,Tsukui:2024} may point to lower than expected turbulence in the population at large \citep{Sheth:2012,Reddish:2022}. 

On the other hand, spiral arms and bars are believed to facilitate the inflow of gas \citep{Shlosman:1989,Englmaier:2000,Englmaier:2004} consistent with observations of disc galaxies in the local Universe \citep{Davies:2009,Yu:2022a,Yu:2022b}. These secular dynamical processes must be operating in the inner sub-kpc to transport gas within the central $\sim100$~pc, thereby fueling nuclear starburst and supermassive black hole growth \citep{Jogee:2006,Angles-Alcazar:2021}. Therefore, we expect secular processes to operate in DSFGs, at least on small scales. While galaxies in the local Universe occasionally show evidence of nuclear spirals (e.g. \citealt{Combes:2013,Audibert:2019}), it remains unknown whether DSFGs exhibit these features within their central kpc. High-resolution imaging ($\lesssim100$~pc) could answer this question and, thus, provide key insights into the structural evolution of DSFGs. At cosmic noon, these scales are accessible only with the aid of gravitational lensing magnification. 

Here, we investigate dust emission from a $z=2.78$ DSFG, lensed by a $z=0.414$ foreground elliptical galaxy \citep{Bothwell:2013}. Following gravitational lens modelling of long-baseline Atacama Large (sub-)Millimetre Array (ALMA) observations reported by \cite{Stacey:2024}, we analyse the reconstructed source and find a nuclear spiral morphology. In Section~\ref{sec:data} we report the source reconstruction methods. In Section~\ref{sec:results}, we report the results of our analysis, and discuss their implications and direction of future work in Section~\ref{sec:disc}. Finally, we summarise our work in Section~\ref{sec:conc}. Throughout, we assume flat $\Lambda$CDM cosmology \citep{PlanckCollaboration:2020}.

\section{Data and lens modelling}
\label{sec:data} 

\begin{figure}
    \includegraphics[width=0.47\textwidth]{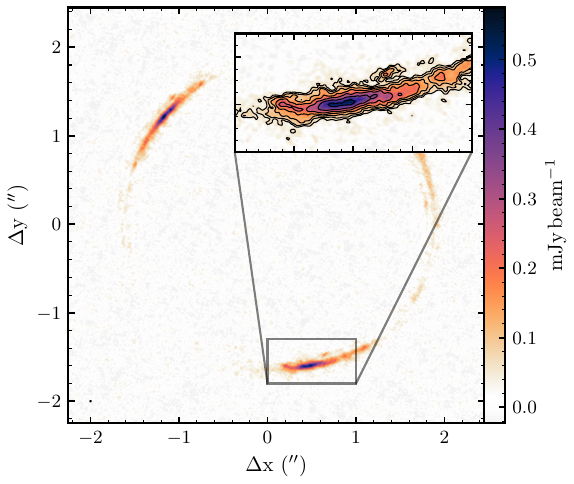} 
    \vspace{-4pt}
    \caption{Image of SPT\,0538$-$50 at 350~GHz. The inset box shows a zoom of the southern image with signal-to-noise ratio contours in steps of 3, $3\sqrt{2}$, 6.. etc. The synthesised beam FWHM is $0.029\times0.026$~arcsec with a position angle 68~deg East of North.
    }
    \label{fig:clean}  
\end{figure}

ALMA observations of SPT\,0538$-$50 (SPT-S\,J053816-5030.8) in band 7 were obtained from the archive associated with project code 2016.1.01374.S (PI: Hezaveh). The continuum-only observations at 350~GHz correspond to the rest-frame 230\um\ in thermal dust emission. The data calibration and reduction were conducted as detailed by \cite{Stacey:2024}. 

The lens modelling was performed by \cite{Stacey:2024} using {\tt pronto}, a semi-linear inversion methodology adapted for interferometric data solving simultaneously for the lens model parameters and source surface brightness (\citealt{Powell:2021,Powell:2022}, but see also \citealt{Vegetti:2009,Rybak:2015a,Rizzo:2018,Ritondale:2019a,Ndiritu:2024}). The lens model comprises an ellipsoidal power-law, external shear and multipole expansions up to 4th order. The source is constructed on a Delaunay grid adapted to the lens model magnification and a regularisation hyper-parameter enforces a correlation between neighbouring points of the source. A `clean' image of the data is shown in Fig.~\ref{fig:clean} made by convolving the maximum a-posteriori sky model with the Gaussian fit to the dirty beam (the synthesised beam) and adding it to an image of the residual visibilities (data$-$model).\footnote{Note that this method of creating an image via forward modelling results in significantly fewer artefacts than with the Clean algorithm.}. 

In the lens modelling procedure, assumptions must be made about the mass distribution in the lens and the smoothness of the source in the form of Bayesian statistical priors. \cite{Stacey:2024} minimised these assumptions by allowing for additional freedom in the projected mass distribution of the lens. However, here, we also consider different types of regularisation when fitting the source surface brightness. As recently discussed by \cite{Galan:2024}, different types of regularisation may be appropriate for different sources, lead to stronger or weaker regularisation of the source surface brightness elements and introduce different biases on lens model parameters. We consider three types of source regularisation: gradient, curvature and area-weighted gradient. Gradient and curvature minimise the gradient and curvature of the source surface brightness, respectively, while area-weighted gradient is a modification of the former whereby the regularisation is weighted by the area of each triangle rather than each triangle being weighted equally. In an extension of the analysis by \cite{Stacey:2024}, we infer {\it maximum a posteriori} lens models and their associated sources for the three regularisation types, casting every pixel into the source plane. Gradient and curvature regularisation types produce model parameters consistent within 2$\sigma$ of those reported in \cite{Stacey:2024}. The area-weighted gradient regularisation model is consistent except for the shear and ellipticity position angles which differ by 2 and 10 degrees, respectively. These discrepancies have a minimal effect on the source as the shear is small ($\approx1$~percent of the convergence) and the lens is very round (axis ratio of $\approx0.9$).

The posterior parameter distribution of the lens parameters inferred by \cite{Stacey:2024} are narrow such that the source surface brightness uncertainty is dominated by noise and artefacts in the data caused by phase and amplitude errors (see also \citealt{Rizzo:2021,Stacey:2021}). We generate uncertainty maps for all three models neglecting the uncertainties in lens model parameters. We inferred an uncertainty in the source plane by generating 1000 mock visibility data sets from the maximum a-posteriori lens model and realisations of the noise and performing linear inversions. We then created mean, standard deviation and signal-to-noise ratio maps for each pixel of the reconstructed source. Additionally, we created source-plane magnification maps by mapping the triangulated image pixels into the source-plane and calculating the ratio of the triangle areas.

\begin{figure*}
\centering
\includegraphics[width=0.905\textwidth]{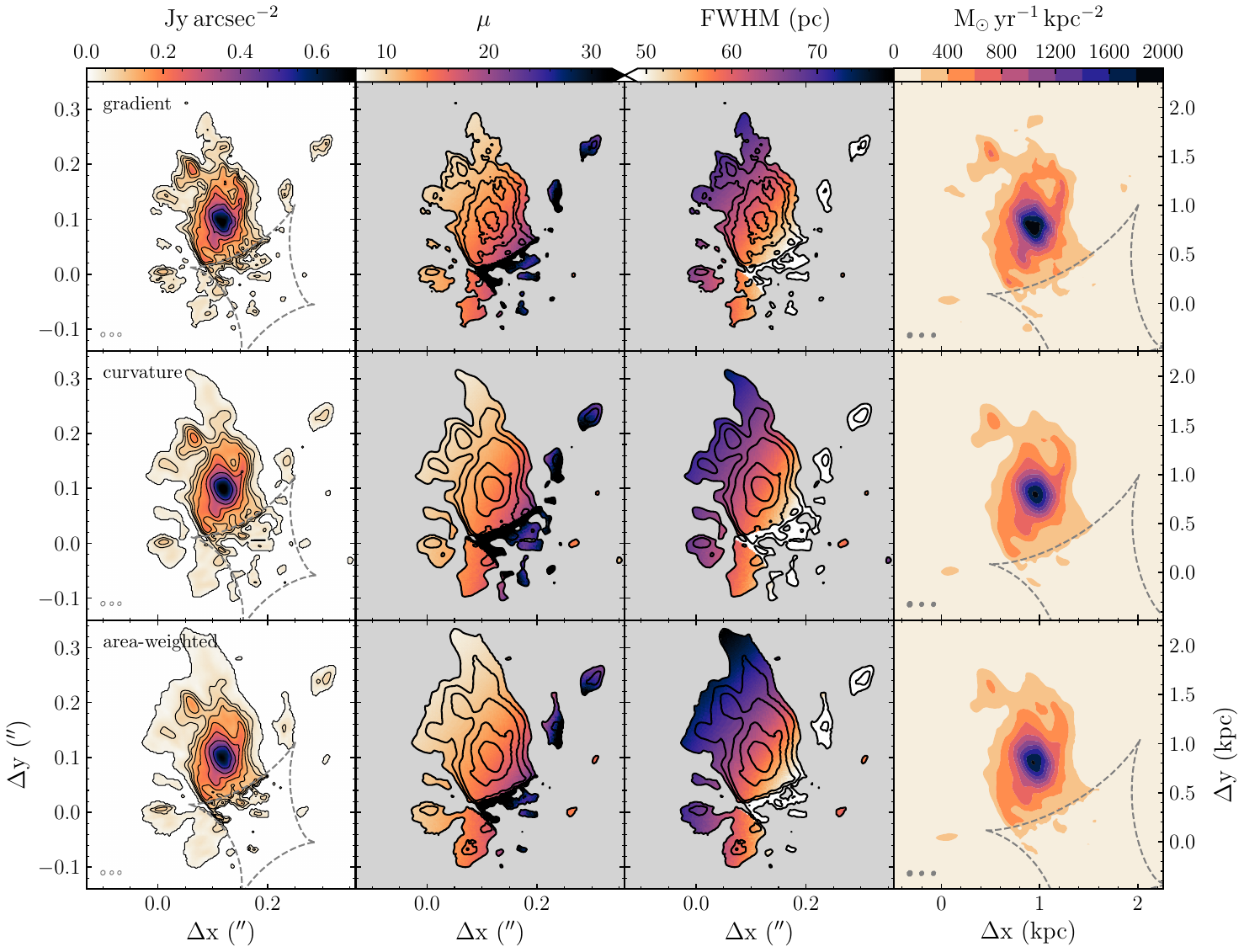}
\vspace{-4pt}
\caption{Reconstructed dust continuum of SPT\,0538$-$50. Top to bottom rows: gradient, curvature and area-weighted gradient regularisation types, all with the same colour scales. First column: surface brightness (area below 3$\sigma$ significance has been masked, where $\sigma$ accounts for the noise, artefacts in the data and the non-linear effects of lensing); contours are in steps of 0.02 Jy\,arcsec$^{2}$ and the dashed grey lines show the tangential caustics. Second column: magnification with contours of signal-to-noise ratio in steps of 3$\sigma$. Third column: the effective magnification-corrected beam FWHM in the source plane with contours of signal-to-noise ratio. Fourth column: star formation rate surface density and caustics in grey. Ellipses in the bottom-left corner show three magnification-corrected beams in the source plane for $\bar{\mu}-\sigma_{\mu}$, $\bar{\mu}$, and $\bar{\mu}+\sigma_{\mu}$, where $\bar{\mu}$ is the mean and $\sigma_{\mu}$ is the standard deviation.}
\label{fig:reconstruction}
\end{figure*}

\section{Results}
\label{sec:results}

As the magnification varies over the source, the effective angular resolution in the source plane is not uniform. We consider two methods to describe the source resolution. Assuming the beam is uniformly magnified, we infer a surface-brightness-weighted mean FWHM of $55$~pc (consistent for all three regularisation types). This is very similar to the median distance between the Delaunay vertices when every pixel is cast back into the source plane; $\approx50$~pc. Therefore, we consider 55~pc a reasonable characterisation of the effective spatial resolution which varies from $<50$ to $80$~pc over the source.

The reconstructed sources are shown in Fig.~\ref{fig:reconstruction} (first panels); the source morphologies are very similar. The total lensed flux density we compute from the product of flux density of the reconstructed source and the magnification map is $0.111\pm0.011$~Jy ($0.111\pm0.011$~Jy; $0.114\pm0.011$~Jy) for gradient (curvature; area-weighted) regularisation methods, including nominal $\pm10$\% flux scale accuracy for ALMA calibration. These are consistent with the measurement of $0.125\pm0.005$~Jy with $\pm15$\% flux scale accuracy quoted by \cite{Reuter:2020} from an unresolved flux density measurement with APEX/LABOCA at the same frequency. This suggests that there is no significant contribution (<10\%) from undetected, extended low surface brightness dust emission. 

Fig.~\ref{fig:reconstruction} (second panels) shows the source-plane magnification for the three reconstructions. Significant variation exists over the source; consequently, the effective beam size varies significantly (Fig.~\ref{fig:reconstruction}; third panels). The magnification is very high at and inside the tangential caustics, so the source appears more granular in this region. Additionally, the low surface brightness emission is more apparent where the magnification is lower, with curvature and area-weighted gradient regularisation types more effectively recovering this emission.

An obscured star formation rate of $700\pm200$~\sfr\ was computed by \cite{Reuter:2020} for SPT\,0538$-$50  by fitting the far-infrared--mm broad-band spectral energy distribution (SED; a modified black-body and \citeauthor{Kroupa:2001} initial mass function) and with a quoted magnification of $20\pm3$ based on lens modelling with observations at lower angular resolution. We find a surface-brightness-weighted mean magnification of $16.6^{+0.3}_{-0.2}$ ($16.5^{+0.2}_{-0.2}$; $15.8^{+0.3}_{-0.2}$) for gradient (curvature; area-weighted) source regularisation types. With these magnification corrections, we infer an obscured star formation rate of $850\pm270$~\sfr\ ($850\pm270$~\sfr\ and $890\pm290$~\sfr) for gradient (curvature; area-weighted); consistent with \cite{Reuter:2020} within the uncertainties. Note that the uncertainties in the star formation rate from SED fitting dominate the overall uncertainties. The star formation rate density for each source reconstruction is shown in Fig.~\ref{fig:reconstruction} (fourth panels), assuming that the dust temperature and emissivity are uniform over the source. The central region of the source has an implied extreme star formation rate of $\approx2000$\sfrd, which is super-Eddington compared to the Eddington limit of nearby ultra-luminous infrared galaxies \citep{Barcos-Munoz:2017}. 

\begin{figure}
% \hspace{2pt}
\includegraphics[width=0.9\linewidth]{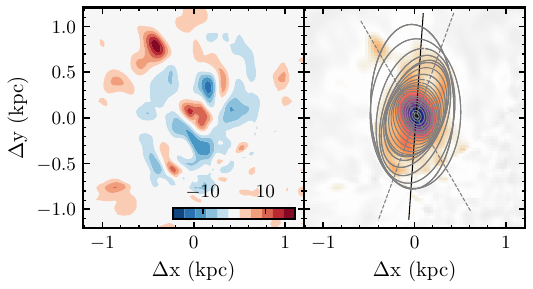}
\vspace{-6pt}
\caption{S\'ersic and isophote fits of SPT\,0538$-$50 (curvature regularisation). Left: noise-normalised residuals of the S\'ersic fit in solid contours with steps of 3. Right: the reconstructed source with isophotes overplotted. The grey dashed lines show the position angle of the isophotes at 0.25 and 0.5 arcsec semi-major axis radius; the black line shows the position angle of the S\'ersic.}
\label{fig:isophotes_overplot} 
\vspace{3pt}
\hspace{5pt}
\includegraphics[width=0.87\linewidth]{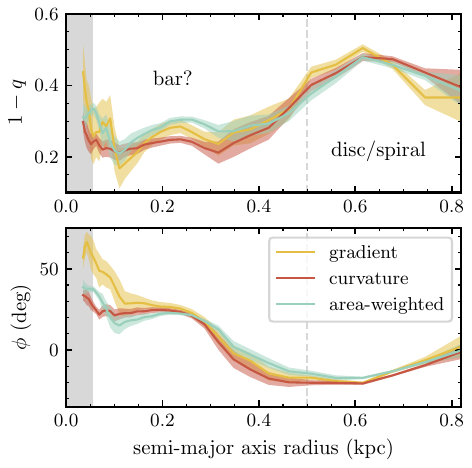}
\vspace{-6pt}
\caption{Ellipticity (top; where $q$ is the axis ratio) and position angle (bottom; defined east of north) of isophotes fitted to the reconstructed source of SPT\,0538$-$50. The shaded coloured regions are the standard deviations of the fits and the grey area shows the median beam FWHM in the source plane. The dashed grey line shows the effective radius of the Sérsic fit. The swing in position angle by 45~deg may be due to a nuclear bar.}
\label{fig:isophotes}
\end{figure}

We fit a 2D S\'ersic profile to the source using a Markov-Chain Monte Carlo sampler ({\sc emcee}; \citealt{emcee}), inferring an effective radius of $0.6\pm0.1$~kpc, S\'ersic index of $1.2\pm0.2$ (where an index of 1 is an exponential disc) and position angle $4\pm3$~deg (East of North). These parameters are consistent with a compact disc, comparable to other lensed DSFGs (e.g. \citealt{Stacey:2021,Rizzo:2021}), but the source structure is not smooth and the S\'ersic provides a poor fit (Fig.~\ref{fig:isophotes_overplot}); the reconstructions for all regularisation types show a dual spiral arm morphology and several clumps with sizes of $\lesssim100$~pc at 3--9$\sigma$ significance. The central region of the source is also not smooth or point-like but resolved into two clumps. These features can also be noted in the lensed images (Fig~\ref{fig:clean}).

To further characterise the source morphology, we fit isophotes using the {\sc isophote} tool in the Python package {\sc photutils} \citep{Bradley:2023}, a technique commonly employed to characterise the structure of barred spirals in the local Universe (e.g. \citealt{Gadotti:2007}). All isophote parameters were free to optimise (centroid, ellipticity, position angle, 3rd and 4th order harmonics). The isophote fit for the source using curvature regularisation is shown in Fig.~\ref{fig:isophotes_overplot}. The ellipticity and position angle of the fitted isophotes for all three regularisation types are shown in Fig.~\ref{fig:isophotes}. The ellipticity position angle swings from $\approx25$~deg at 0.2~kpc to $-20$~deg at 0.6~kpc, with accompanying peaks in ellipticity (labelled in Fig.~\ref{fig:isophotes}), which is a typical signature of a bar (e.g. \citealt{Tsukui:2024,LeConte:2024}).

\section{Discussion}
\label{sec:disc}

Nuclear bars and spirals have occasionally been found in active and inactive galaxies in the nearby Universe \citep{Buta:1993,Martini:2003a,Erwin:2015}) and are predicted by both idealised and hydrodynamical simulations \citep{Shlosman:1989,Englmaier:2004,Angles-Alcazar:2021}. They are found in discs, are produced by an inflow of cold gas, contain young stellar populations, are rotationally supported, and are thought to probe secular dynamical processes involved in the growth of stellar spheroids and AGN fuelling \citep{Knapen:1995,Garcia-Burillo:2005,Combes:2013,Prieto:2019,Audibert:2019,Gadotti:2020}. The effective spatial resolution we have obtained here suggests the presence of spiral arms and a bar on much smaller scales than previously found in DSFGs, yet comparable to the size of observed and simulated nuclear discs (e.g. \citealt{Combes:2013,Audibert:2019,Angles-Alcazar:2021}) in galaxies with similar stellar mass. The disc is $5\to10$ times smaller than the typical $\sim5$~kpc effective radius of the stellar discs of DSFGs \citep{Hodge:2024,Gillman:2024}. Furthermore, the characteristic exponential profile, spiral arms and possible bar suggest that SPT\,0538$-$50 may be dynamically cold, like local nuclear discs \citep{Gadotti:2020}. 

A high gas density is required to explain the apparently super-Eddington star formation rate density in the central $<200$~pc of SPT\,0538$-$50. Indeed, SPT\,0538$-$50 is known to be rich in molecular gas ($\sim10^{10.2}$\msol; \citealt{Bothwell:2013}). The final stages of a gas-rich major merger could result in gas build-up in the central kpc \citep{Jogee:2006}, or potentially an inward migration of gas clumps \citep{Dekel:2014}. However, it seems to be difficult to produce sufficient loss of angular momentum to pile a high density of gas within $\sim100$~pc without the additional influence of secular dynamical processes \citep{Jogee:2006,Angles-Alcazar:2021}. The nuclear spiral arms and bar we see here could facilitate this gas inflow.

Spiral arms can be transient or long-lived: transient two-arm spirals reaching into the centre of the galaxy may be induced by tidal interactions (e.g. \citealt{Dobbs:2010}) or they could be long-lived and produced secularly, such as via swing amplification \citep{Toomre:1981}. Greater clarity would be achieved with observations at matched angular resolution of stars and emission line kinematics. The resolved gas velocity dispersion, in combination with the resolved stellar distribution, star formation and gas density, could test models of turbulence associated with secular dynamics or gravitational instabilities induced by mergers/interactions \citep{Rizzo:2024}. Resolved kinematics on these scales may also reveal evidence for non-circular motions along the spiral arms or bar and test whether this region is kinematically distinct from the main galaxy disc (e.g. \citealt{Davies:2009,Lelli:2022,Roman-Oliveira:2023,Tsukui:2024}). Additionally, it could be that the spiral morphology of SPT\,0538$-$50 represents only the centre of spiral arms that extend out to several kpc in unobscured stellar emission, as frequently seen for high-redshift galaxies with JWST \citep{Kuhn:2023,Hodge:2024,Gillman:2024}. Furthermore, nuclear discs in the local Universe are usually found within bars \citep{Gadotti:2020}: follow-up observations in the rest-frame optical/infrared will determine this for SPT\,0538$-$50.

Another consideration with SPT\,0538$-$50 is whether the compact central dust emission could be attributed to heating by an AGN. Simulations find that a strong AGN radiation field could lead to higher effective dust temperature and integrated star formation rate inferred from broad-band SED fitting \citep{DiMascia:2023}. When computing the reconstructed star formation rate density (Fig.~\ref{fig:reconstruction}; fourth panel), we assumed that the dust emissivity and temperature are uniform over the source, but in reality, these are likely non-uniform. A radially decreasing temperature gradient (e.g. \citealt{Walter:2022}) would imply an even higher central density of star formation than we inferred here. Multi-frequency dust continuum observations at matched angular resolution would be required to test whether there is such a temperature gradient and compute the star formation rate density more robustly. While there is no clear evidence for an AGN in the broad-band SED for this object \citep{Bothwell:2013}, it also cannot be ruled out. X-ray or radio observations (currently lacking) are warranted to test for the presence of radio jets or an accretion disc corona which definitively characterise AGN activity. 

Future work should involve the analysis of a sample of lensed DSFGs observed at similar angular resolution with ALMA to determine whether nuclear spirals are prevalent in this population. Crucially, we find that smoothing the source to the native resolution of the observations (30~mas) erases any evidence of the spiral morphology, suggesting that gravitational lensing is the only feasible approach to resolving these features. A prevalence of nuclear discs and bars in the DSFG population would indicate these physical mechanisms help drive their intense star formation and mass growth.

Finally, a common concern about studying strongly lensed sources is that the results depend on a lens model that involves strong assumptions and poorly understood uncertainties. Here, our source error analysis gives due diligence to questions of source structure significance, such as whether individual clumps are genuine or noise artefacts. We find that changes in the regularisation type and additional lens model freedom in the form of angular structure do not significantly alter the inferred source properties, in agreement with the recent study with mock optical data by \citep{Galan:2024}. Furthermore, \cite{Stacey:2024} found that the inclusion of angular structure in the lens model did not change the source structure compared to the canonical power-law plus external shear. While these results suggest that our findings are robust to lens modelling systematics, our analysis does not consider the effect of a mass sheet transform-like effect that may result if the lens is part of a group. This would reduce or inflate the size of the lensed source, along with its inferred star formation rate, without changing its structure (e.g. see the effect of including the foreground in Fig. 7 of \citealt{Powell:2022}). It is unknown whether the lens of SPT\,0538$-$50 is part of a group \citep{Stacey:2024}, so we leave such investigation to future work.

\section{Conclusions}
\label{sec:conc} 

We have shown how sophisticated gravitational lens modelling tools and long-baseline ALMA observations can be combined to produce observations with spatial resolution sufficient to resolve the inner structure of DSFGs. For SPT\,0538$-$50, we find the central kpc of the galaxy exhibits a dual spiral arm morphology and a potential nuclear bar that could facilitate gas inflow, feeding the nuclear starburst and supermassive black hole. These results show that our approach can probe the physical drivers of spheroid formation at cosmologically interesting epochs. More definitive answers on the role of nuclear spirals in the structural evolution of DSFGs can only come from gas kinematics. Strong gravitational lensing magnification is the only feasible way to reach these physical scales. Here, we have demonstrated that the lensed source reconstructions are robust to a combination of angular flexibility in the lens model and different choices of source regularisation. We found that the factors limiting the overall uncertainties on the lensed source properties are instrumental flux calibration and spectral energy distribution modelling.

\begin{acknowledgements}
Our analysis made use of NumPy, SciPy, Matplotlib, Astropy and Photutils packages for Python \citep{Harris:2020,Virtanen:2020,Hunter:2007,Astropy:2013,Astropy:2018,Bradley:2023}. We used ALMA data associated with project code 2016.1.01374.S. ALMA is a partnership of ESO (representing its member states), NSF (USA) and NINS (Japan), together with NRC (Canada), MOST and ASIAA (Taiwan), and KASI (Republic of Korea), in cooperation with the Republic of Chile. The Joint ALMA Observatory is operated by ESO, AUI/NRAO and NAOJ. This work is based on research supported in part by the National Research Foundation of South Africa (Grant Number: 128943).
\end{acknowledgements}

%%%%%%%%%%%%%%%%%%%% REFERENCES %%%%%%%%%%%%%%%%%%

\bibliographystyle{aa_url}
\bibliography{references} 

\end{document}